\documentclass[acus]{JAC2003}

\usepackage{graphicx}
\usepackage{booktabs}

\begin{document}
\title{TE WAVE MEASUREMENT AND MODELING\thanks{This work is supported by the US National Science Foundation PHY-0734867, PHY-1002467 and the US Department of Energy DE-FC02-08ER41538, DE-SC0006505.
}}

\author{J.P. Sikora\thanks{ jps13@cornell.edu}, R.M. Schwartz, K.G. Sonnad, CLASSE, Ithaca, New York 14853 USA\\
D. Alesini, INFN/LNF, Frascati (Roma)\\
S. De Santis, LBNL, Berkeley, California, USA}

\maketitle

\begin{abstract}

In the TE wave method, microwaves are coupled into the beam-pipe and the effect of the electron cloud on these microwaves is measured.  An electron cloud (EC) density can then be calculated from this measurement. There are two analysis methods currently in use. The first treats the microwaves as being transmitted from one point to another in the accelerator. The second more recent method, treats the beam-pipe as a resonant cavity. This paper will summarize the reasons for adopting the resonant TE wave analysis as well as give examples from C{\small ESR}TA and DA$\Phi$NE of resonant beam-pipe. The results of bead-pull bench measurements will show some possible standing wave patterns, including a cutoff mode (evanescent) where the field decreases exponentially with distance from the drive point. We will outline other recent developments in the TE wave method including VORPAL simulations of microwave resonances, as well as the simulation of transmission in the presence of both an electron cloud and magnetic fields.

\end{abstract}

\section{INTRODUCTION}

Beam Position Monitor (BPM) buttons are a convenient way to couple microwaves in and out of the beam-pipe. In most of the measurements described here, a fixed frequency is used for excitation and the response of the system is measured with a spectrum analyzer. Figure~\ref{EC12_sikora:trans} shows a typical setup, where a circulator is used to protect the amplifier from the beam induced signal and a band-pass filter to limit the beam induced signal at the input of the spectrum analyzer. Although not shown in the figure, connections are often made to pairs of buttons through a hybrid combiner, so that the direct beam signal is mostly canceled and the desired TE wave excitation or reception is maximized. 
\begin{figure}[htb]
   \centering
   \includegraphics*[width=65mm]{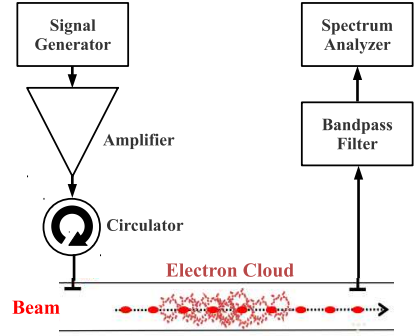}
   \caption{Basic setup for TE wave measurements, where microwaves are coupled in/out of the beam-pipe using Beam Position Monitor (BPM) buttons.}
   \label{EC12_sikoratrans}
\end{figure}

\section{TRANSMISSION VS. RESONANCE}

There are two principal variations of the TE wave method for the measurement of EC density. The first is the original transmission method, where TE waves are propagated through a beam-pipe as if it were a waveguide. In this method, a fixed frequency TE wave will be phase shifted by the presence of an electron cloud between the input and output points~[1-4]. The magnitude of the phase shift in the absence of a magnetic field is given by Eq.~\ref{EC12_sikora:eq_pm_trans}, where $\omega$ is the angular frequency of the transmitted signal, $\omega_c$ the cutoff frequency of the waveguide, $n_e$ is EC density and $L$ the distance between the input and output points along the beam-pipe~\cite{PAC07:THPAS008}. If the EC density varies periodically (as happens with a short train of bunches in a storage ring), the result will be a phase modulation of the original signal.

\begin{equation}
\Delta \phi \; = \;
 \frac{ L \; e^2 n_e }{2 m_e \varepsilon_0 c(\omega^2 - \omega_c^2)^{1/2} } .
	\label{EC12_sikora:eq_pm_trans}
\end{equation}

This method requires long sections of beam-pipe of uniform cross section, since changes in the geometry will produce reflections. These reflections will make the propagation length $L$ difficult to quantify, since the microwaves will make more than one transit through the electron cloud and will generally extend beyond the distance $L$.

\begin{figure}[ht]
   \centering
   \includegraphics*[width=65mm]{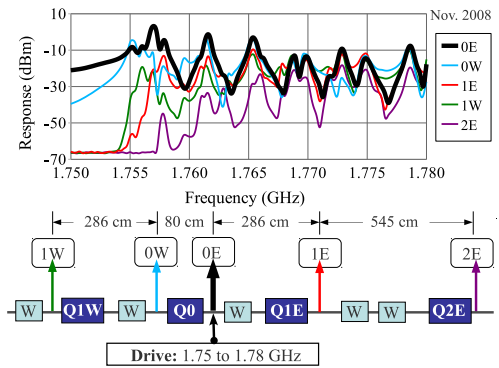}
   \caption{This is the measured response when exciting beam-pipe with TE waves at 0E and receiving at other detectors in the region. The wide trace is the response when driving and receiving at 0E.}
   \label{EC12_sikora:L0_Response}
\end{figure}

The need for analysis based on resonance came from initial TE wave measurements that were made at C{\small ESR}TA in 2008. A new configuration of beam-pipe was installed in the L0 section of C{\small ESR}TA that included a number of beam-pipe transitions, instrumentation, longitudinal slots to connect ion pumps to the vacuum space and six superconducting wigglers. Measurements were made of the transmission of microwaves from BPM 0E near the middle of this section to several other points of the assembly, including both exciting and receiving at 0E. Fig.~\ref{EC12_sikora:L0_Response} shows an overlay of these measurements. 

\begin{figure}[ht]
   \centering
   \includegraphics*[width=65mm]{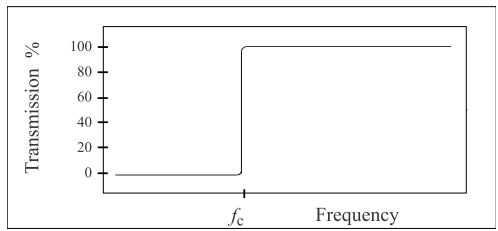}
   \caption{
The response of an ideal waveguide is shown schematically, where there is no propagation through the waveguide below its cutoff frequency $f_c$ and free propagation of the microwaves above $f_c$.}
   \label{EC12_sikora:wg_response}
\end{figure}

The measured response in Fig.~\ref{EC12_sikora:L0_Response} bears little resemblance to ideal waveguide response (see Fig.~\ref{EC12_sikora:wg_response}), except that it generally increases at higher frequencies. The dominant feature of Fig.~\ref{EC12_sikora:L0_Response} is the series of resonances that can be seen over this frequency range. Also, many of these resonances occur at the same frequencies whether the location of the receiver is at the drive point or elsewhere in this section of beam-pipe. For example, near 1.775~GHz all of the detectors show a resonant peak and all have similar amplitudes. This suggests the excitation of a single resonator with different choices of the location of the pickup.

\begin{figure}[ht]
   \centering
   \includegraphics*[width=65mm]{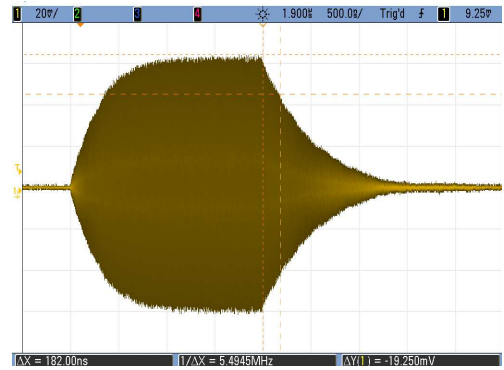}
   \caption{This scope trace was obtained by exciting and receiving at BPM 0W and shows a cavity damping time of about 500~ns corresponding to a Q of 3000.}
   \label{EC12_sikora:damping}
\end{figure}

Following the idea that the microwaves were exciting resonances within the beam-pipe, a pulsed source of RF was used to drive one of the resonances for 2~$\mu$s, so that the growth and decay of the field could be observed. The resulting 500~ns damping time is shown in Fig.~\ref{EC12_sikora:damping}

\subsection{Examples of Resonant Beam-pipe}

\begin{figure}[ht]                           

  \begin{minipage}[t]{.48\textwidth}
    \begin{center}
    \includegraphics*[width=65mm]{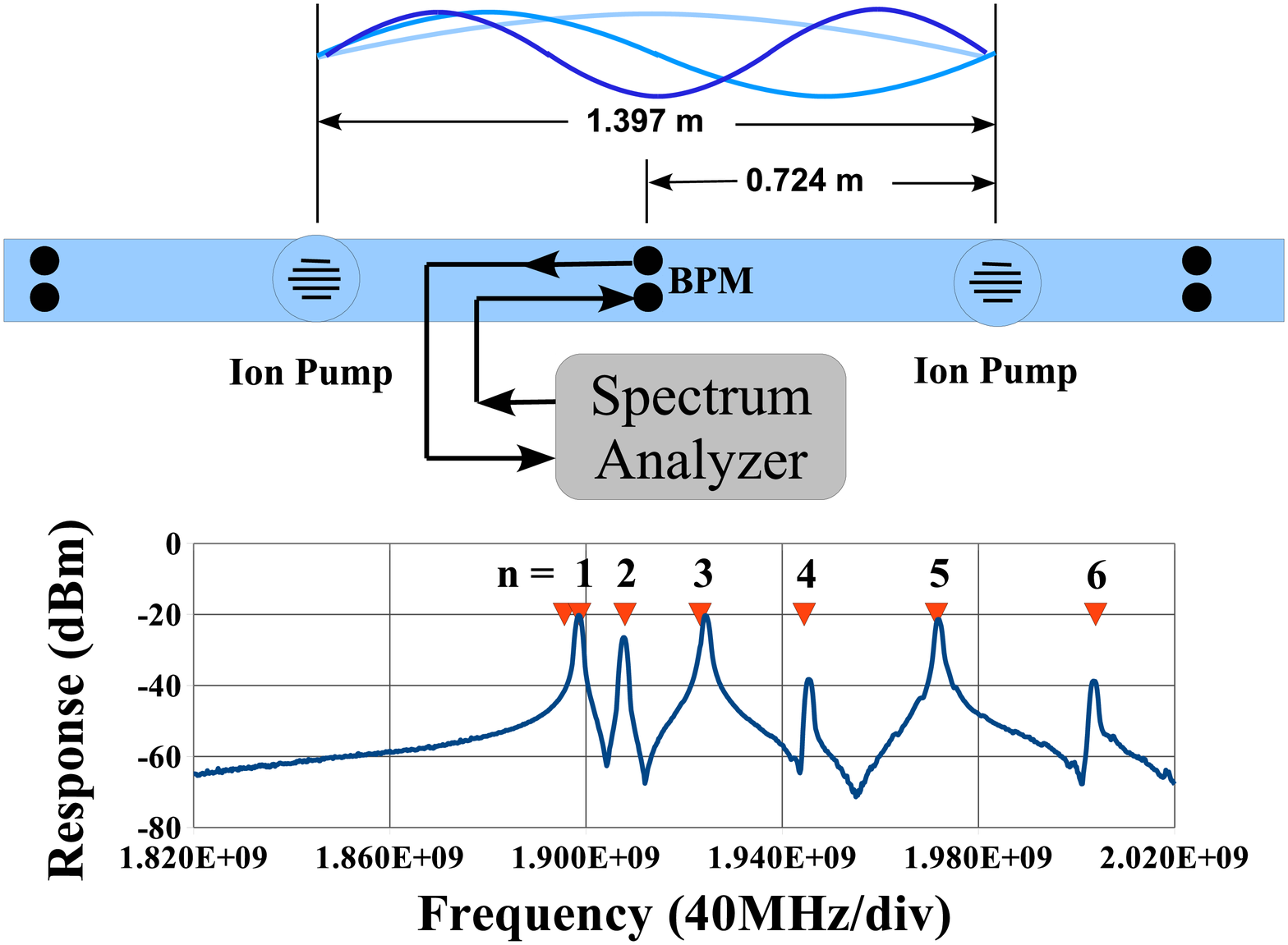}
    \caption{The response at 43E is very similar to that of a rectangular cavity. The numbered triangles in the plot correspond to the values of n of Eq.~\ref{EC12_sikora:eq_rectang}. }
    \label{EC12_sikora:43E}
    \end{center}
  \end{minipage}

  \begin{minipage}[t]{.48\textwidth}
    \begin{center}
    \includegraphics*[width=65mm]{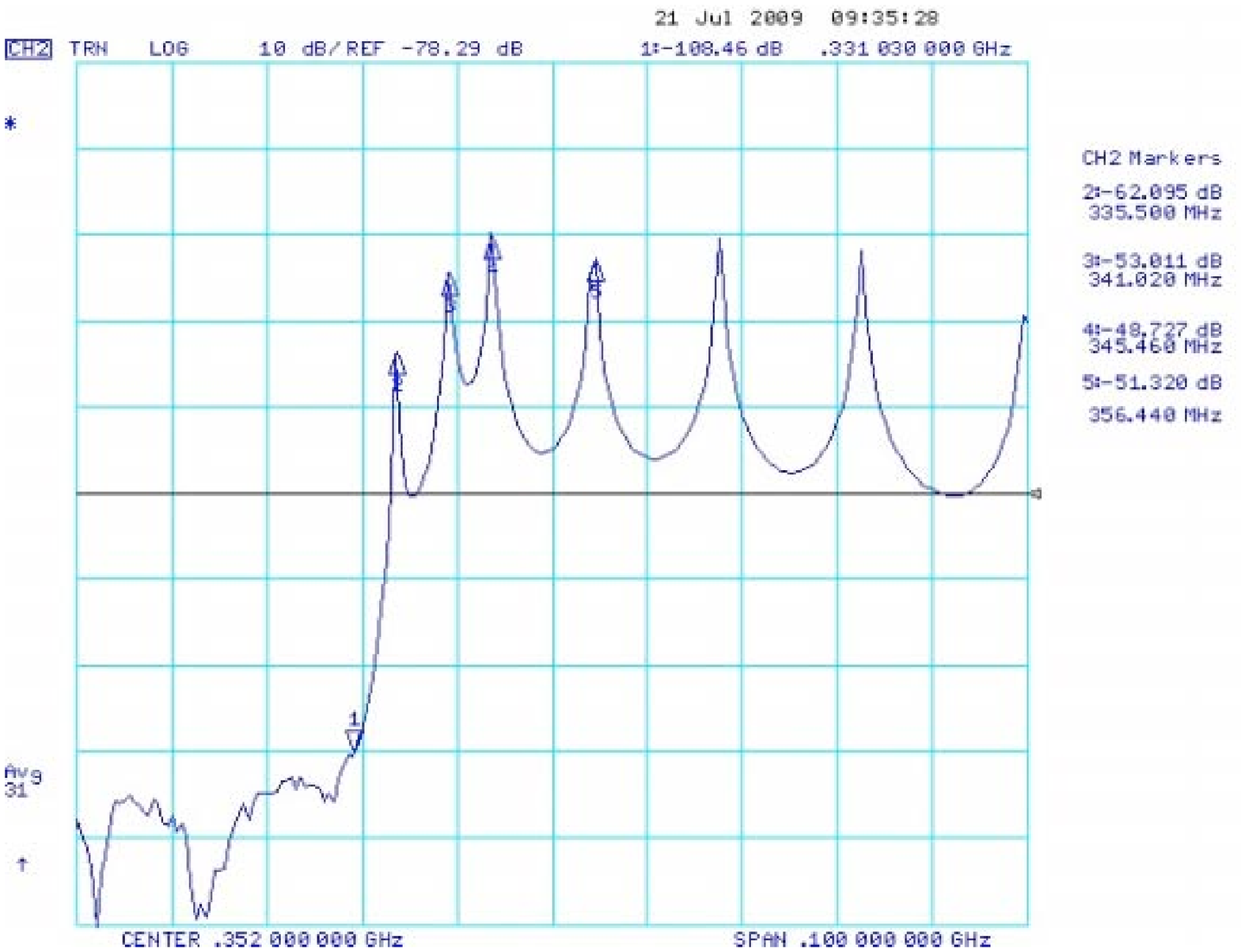}
    \caption{This response from the Advance Photon Source (Argonne) also matches that of a rectangular cavity and is produced by reflections at bellow end flanges.}
    \label{EC12_sikora:aps}
    \end{center}
  \end{minipage}
\end{figure}

Since the initial observations of the L0 section of C{\small ESR}TA, other measurements that suggest resonance have been made at C{\small ESR}TA and at other laboratories. A relatively simple example was obtained at 43E in C{\small ESR}TA. At that location, BPMs are located in a smooth straight section of beam-pipe between two ion pumps as shown in Fig.~\ref{EC12_sikora:43E}. At the ion pumps, longitudinal slots connect the pump volume to that of the beam-pipe. These slots reduce the coupling of beam-induced signal into the pumps, but also generate reflections for a TE wave that is propagating through the beam-pipe. So when excited at the BPM between the pumps, standing waves are set up between the pumps that are similar to those of a rectangular cavity, whose resonant frequencies are given in Eq.~\ref{EC12_sikora:eq_rectang}.

\begin{equation}
    f^2 \;  = \; f_{c}^{2} + \left(\frac{nc}{2L} \right)^2
	\label{EC12_sikora:eq_rectang}
\end{equation}

Another example was observed at the Advanced Photon Source (APS) at Argonne National Laboratory. In this case, the resonances were not used for electron cloud measurements, but were sometimes excited by the beam. This somewhat unpredictable excitation gave varying offsets in the beam position monitor system~\cite{BIW10:TUPSM049}. The reflections were set up between bellow end flanges in the dipole sections with the response shown in Fig.~\ref{EC12_sikora:aps}.

\begin{figure}

  \begin{minipage}[t]{.48\textwidth}
    \begin{center}
    \includegraphics*[width=65mm]{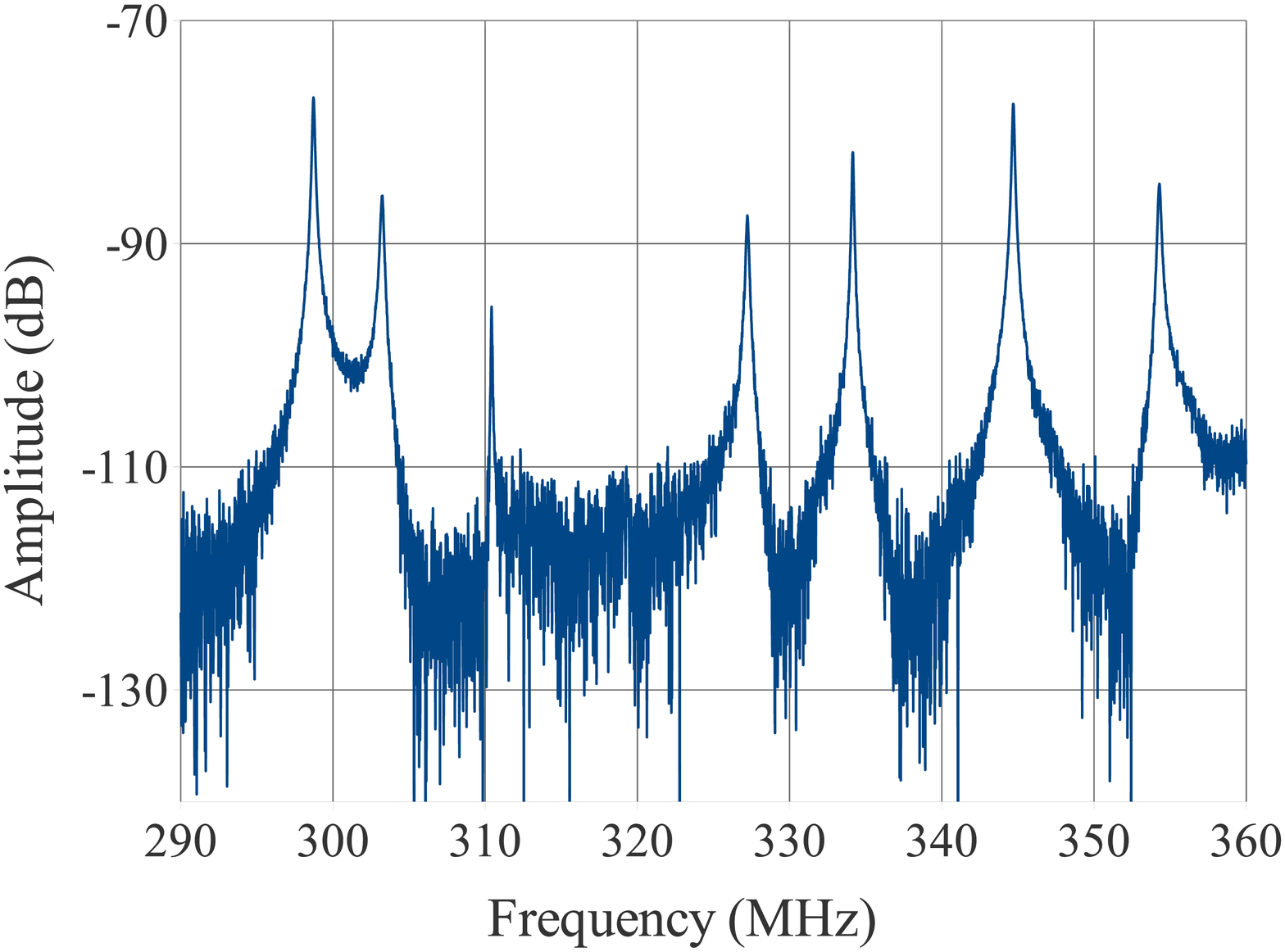}
    \caption{A response measured in the positron storage ring at DA$\Phi$NE shows a number of resonances, although they are not consistent with a simple rectangular cavity.}
    \label{EC12_sikora:dafne_res}
     \end{center}
  \end{minipage}

  \begin{minipage}[t]{.48\textwidth}
    \begin{center}
    \includegraphics*[width=65mm]{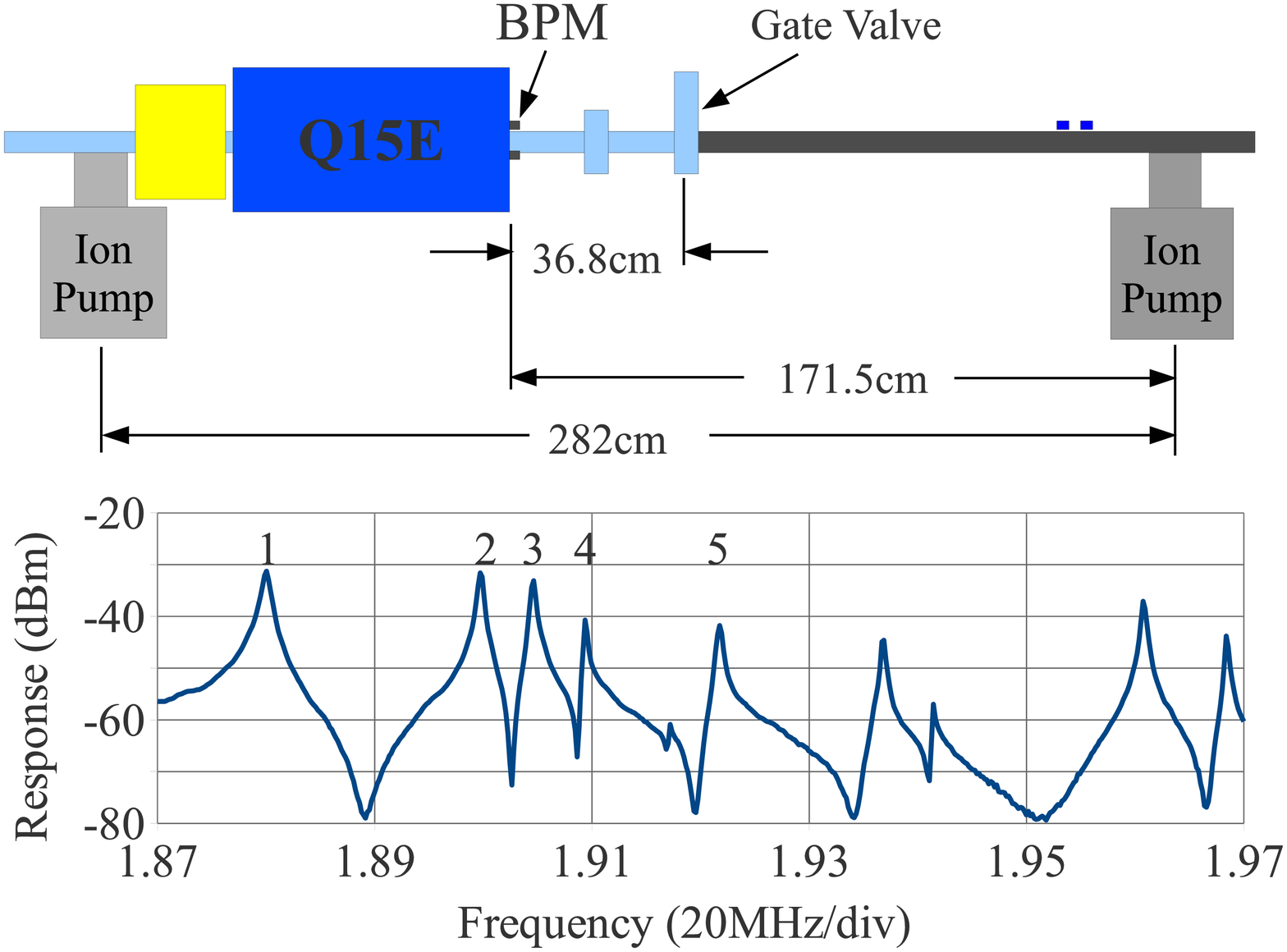}
    \caption{The response measured at 15E in C{\small ESR}TA is similar to a rectangular cavity, except that the lowest resonance is much lower in frequency than would be expected.}
    \label{EC12_sikora:15E}
    \end{center}
  \end{minipage}

\end{figure}

Resonances have also been observed in the positron storage ring at DA$\Phi$NE~\cite{IPAC12:Alesini}. The sequence of resonances shown in Fig.~\ref{EC12_sikora:dafne_res} are not very well matched by a simple rectangular cavity and further study would be needed to understand this response. 

At C{\small ESR}TA, 15E is another location that has a relatively simple geometry, with a BPM between two ion pumps (having longitudinal slots). The response at 15E does not match that of a rectangular cavity as well as 43E. This is probably due to a gate valve near the midpoint of the resonant section of 15E. The lowest resonance is considerably lower than the others as can be seen in  Fig.~\ref{EC12_sikora:15E}.  It was  relatively easy to run low loss cables to and from this location so that data could be taken with beam. 

\subsection{EC Density Calculation for Resonances}

TE wave resonance is an alternative analysis method that makes use of the reflections that are commonly found in beam-pipe. This method treats the beam-pipe and its reflections as a resonant cavity. The presence of an electron cloud will shift the frequencies of the resonances and produce a corresponding effect on the observed signal.

Fortunately, the use of cavities for the measurement of plasma densities is reasonably well understood~\cite{MAHeald1965:PlasDiagMicroW}. In plasma physics, densities are measured that are many orders of magnitude higher than the EC densities found in accelerators. While an EC density of $10^{12}$ per cubic meter would be considered high in an accelerator, plasma physicists would be dealing with a similar density per cubic \textit{centimeter}. Perturbation techniques can be used to calculate the change in the cavity's resonant frequency vs. plasma (EC) density~\cite{MAHeald1965:PlasDiagMicroW,PR106:196,IEEE:Carter}. The result is especially simple for low densities where the collision frequency is essentially zero and in the absence of an external magnetic field. In this case the frequency shift is given by Eq.~\ref{EC12_sikora:Ne}.

\begin{equation}
\frac{\Delta \omega}{ \omega} \; = \;
 \frac{e^{2} }{2 \varepsilon _{0} m_{e} \omega ^{2} } \frac{ \displaystyle \int_{V} n_{e} E_{0}^{2}\,dV  }
       { \displaystyle \int_{V} E_{0}^{2} \,dV } .
	\label{EC12_sikora:Ne}
\end{equation}

Where $n_e$ is the local EC density, $\omega$ is the cavity frequency, $E_0$ is the local magnitude of the rf electric field and the integrals are over the cavity volume $V$. 
 
So if the beam-pipe is considered to be a resonant cavity and the electron cloud a plasma within its volume, it can be straightforward to calculate the relation between the EC density $n_e$ and the frequency shift~\cite{IPAC11:SIKORA_TEW}. For example, if the EC density is assumed to be uniform over the volume of the cavity, this constant $n_e$ can be taken outside of the integral of Eq.~\ref{EC12_sikora:Ne} and the two integrals will be identical -- cancelling each other. The electron cloud density $n_e$ can then be calculated directly from the frequency shift $\Delta \omega / \omega$ using physical constants. Where the EC density is not uniform, the upper integral gives an average of the local density $n_e$ that is weighted by the local $E_0^2$ over the cavity volume $V$.

In some circumstances it is possible to measure the frequency shift directly, as when a storage ring is almost entirely filled with bunches -- generating an EC density that is nearly constant with time. For bunch trains and electron cloud lifetimes that are short compared to the storage ring revolution period, there is a modulation of the EC density and a corresponding modulation of the cavity's resonant frequency. 

The effect of an EC density modulation on the measured signal is generally more complicated than that obtained by transmission analysis. If the duration of the cloud is similar to the damping time of the resonance, the transient response becomes important. The effect of the modulated resonant frequency with a fixed drive frequency is a combination of frequency, phase and some amplitude modulation. The relative weight of the three types of modulation depends on the electron cloud duration compared to the beam revolution time as well as the damping time of the resonance~\cite{IPAC12:DeSantis}.

\subsection{Examples of Resonance Data}

\begin{figure}[ht]
   \centering
   \includegraphics*[width=65mm]{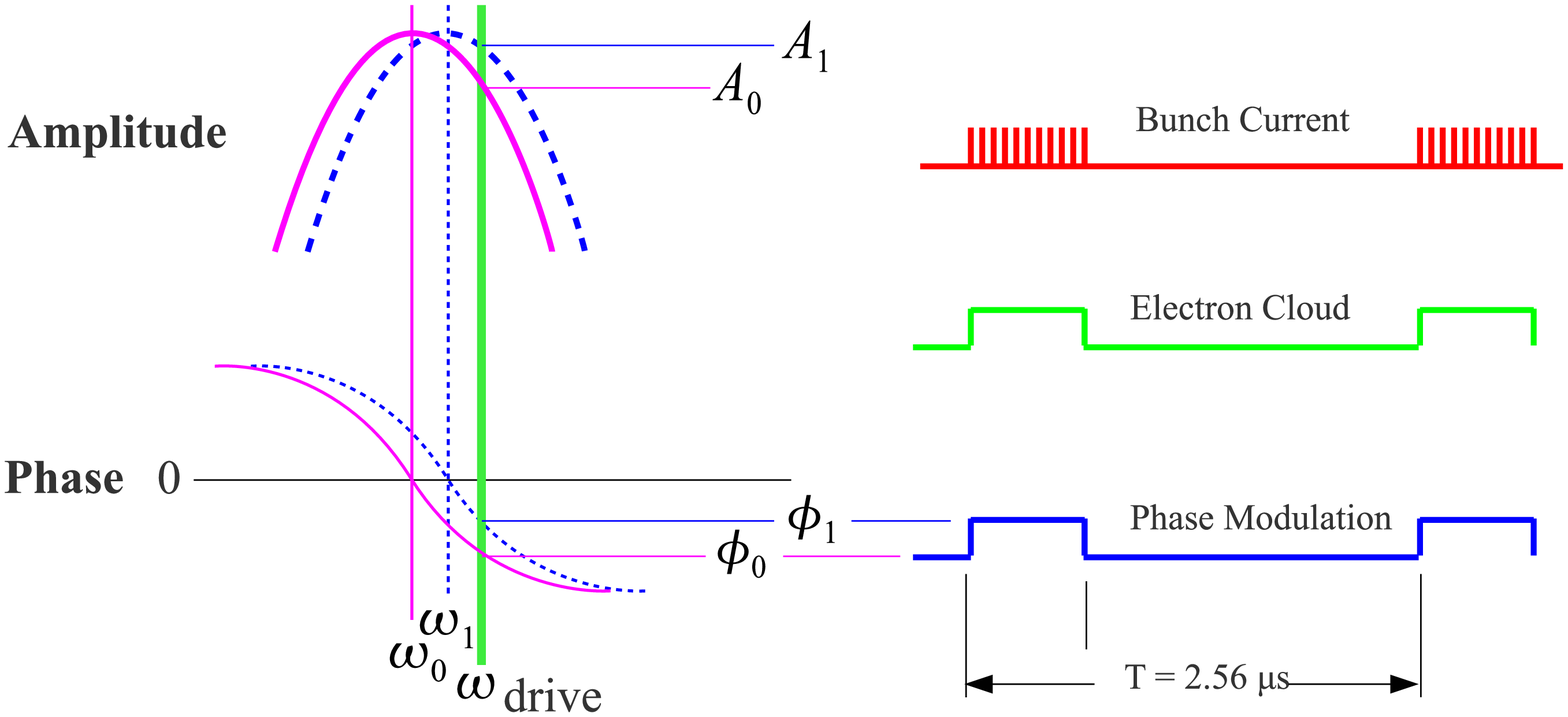}
   \caption{A fixed frequency is used to drive the beam-pipe resonance. If the damping time of the resonance is short compared to the duration of the cloud, the modulating EC density will produce a change in both the phase and amplitude of the response. }
   \label{EC12_sikora:phase_mod}
\end{figure}

In generating plots of TE wave resonant data from C{\small ESR}TA, some simplifying assumptions have been made: the EC density has a fixed value for the duration of the bunch train and is zero otherwise; the EC duration is long compared with the damping time of the TE wave resonance; and the effect of magnetic fields can be neglected. With these assumptions, the effect of the shift in resonant frequency is shown in Fig.~\ref{EC12_sikora:phase_mod}. With a fixed drive frequency, both the phase and amplitude of the cavity response will change with resonant frequency. The amplitude change is minimized when driven close to the resonant frequency. 

For small modulations near resonance, the expressions for the amplitude and phase shifts are given in Eq.~\ref{EC12_sikora:phi}.

\begin{eqnarray}
x(t)    & = & A_n  \sin(\omega t + \phi_n )\\
\nonumber \\
A_n     & = & Q \frac{A}{[ Q^2 (\omega_n^2 - \omega^2)^2 +  \omega^4 ]^{1/2} }  \\
\phi_n & = & \tan^{-1} \left[ Q \frac{ (\omega_n^2 - \omega^2) }{ \omega^2 } \right]
 \label{EC12_sikora:phi}
\end{eqnarray}

\begin{figure}[ht]
   \centering
   \includegraphics*[width=65mm]{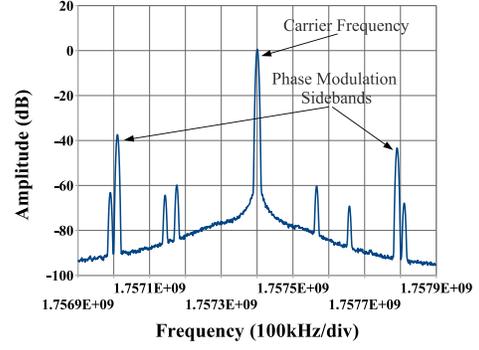}
   \caption{ With a revolution frequency 390~kHz, phase modulation sidebands are visible in the spectrum to either side of the carrier. This data was taken at 15E in C{\small ESR}TA with a 20 bunch train of positrons in the storage ring, 14~ns spacing and about 100~mA total current. }
   \label{EC12_sikora:sidebands}
\end{figure}

With a single bunch train in the C{\small ESR}TA storage ring with a revolution frequency of 390~kHz, the modulated EC density results mostly in phase modulation sidebands 390~kHz away from the carrier frequency as shown in Fig.~\ref{EC12_sikora:sidebands}. In addition to the fixed carrier frequency, the drive signal includes a reference phase modulation of 1~mrad at 410~kHz with sidebands that are also visible in the figure.

\begin{figure}[t]
   \centering
   \includegraphics*[width=65mm]{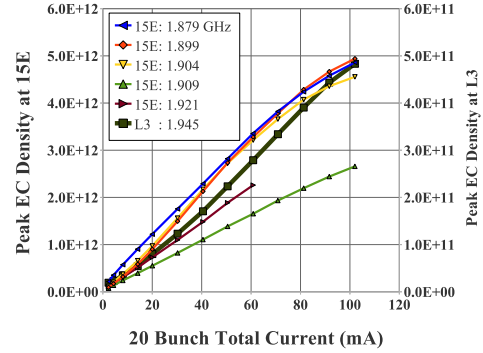}
   \caption{ The EC density calculated from TE wave resonances is plotted vs. total current in a 20 bunch train of positrons. Data from the first five resonances at 15E (Fig.~\ref{EC12_sikora:15E})  was used, as well as data from the L3 region (right-hand scale) where the EC density is much lower. }
   \label{EC12_sikora:tew_15E}
\end{figure}  

To calculate the EC density that produced the measured phase modulation, we make use of the simplifying assumptions made above. In addition, for the expected EC densities of $10^{12}e/$m$^{3}$ and a Q of about 3000, the change in frequency and the resulting phase modulation are small and can be obtained using Eq.~\ref{EC12_sikora:phi}.

With sinusoidal modulation, the ratio of the linear amplitude of the sideband to that of the carrier should be about $\frac{1}{2} \Delta \phi $. But since the modulation is taken to be a rectangular pulse rather than sinusoidal, a Fourier expansion of that pulse should be used to determine the resulting amplitude of the first sideband. 

The data plotted in Fig.~\ref{EC12_sikora:tew_15E} is the result of all of these approximations. Most of this data was taken at 15E in C{\small ESR}TA, with a 20 bunch train of positrons in the storage ring, 14~ns spacing and about 100~mA total current. The data of the wide black plot was taken at the same time, but came from the L3 region of C{\small ESR}TA where the amount of synchrotron light (and the resulting electron cloud) is much lower than at 15E.

\begin{figure}[ht]
   \centering
   \includegraphics*[width=65mm]{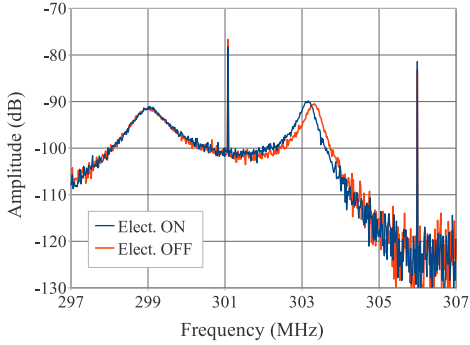}
   \caption{The change in the response of two TE wave resonances in the DA$\Phi$NE positron ring as the clearing electrodes are turned OFF and ON. }
   \label{EC12_sikora:dafne_shift}
\end{figure}

At DA$\Phi$NE, the TE wave resonant frequency shift could be measured directly. Clearing electrodes have been installed in much of the  positron storage ring~\cite{IPAC12:Alesini}. With a positive voltage applied, it is expected that most of the electron cloud would be suppressed at the locations of the clearing electrodes. Fig.~\ref{EC12_sikora:dafne_shift} shows the change in the TE wave resonance response as these clearing electrodes are turned OFF and ON. This data was taken with 800~mA in the positron ring with 100 bunches filled and a 20 bunch gap. This large duty factor allows the resonant frequency shift to be observed directly. Rather than exciting with a fixed frequency, the drive frequency was swept so that the resonant response could be observed. The right peak of  Fig.~\ref{EC12_sikora:dafne_shift} changes by about 200~kHz, corresponding to a change in EC density of $1.5 \times 10^{12}$m$^{-3}$. The fact that the left peak does not shift at all suggests that the standing wave of this resonance is in a region without clearing electrodes. Further analysis is needed in order to understand this variation in the frequency shifts.

\section{BEAD PULL MEASUREMENTS}

As shown in Eq.~\ref{EC12_sikora:Ne}, the sensitivity of the measurement is given by an integral of product of the local EC density $n_e$ and the local $E^2$ of the resonance. Generally, the EC density will not be uniform over the volume (as suggested by the DA$\Phi$NE data) so that the correct interpretation of the data depends on a good understanding of the standing wave pattern that is generated by each resonance. Changes in beam-pipe geometry can result in resonances that differ considerably from a simple rectangular cavity. So it is useful to estimate this standing wave pattern in the beam-pipe, either through physical modeling or by simulation. 

\begin{figure}[ht]
   \begin{minipage}[th]{.48\textwidth}
   \begin{center}
   \includegraphics*[width=65mm]{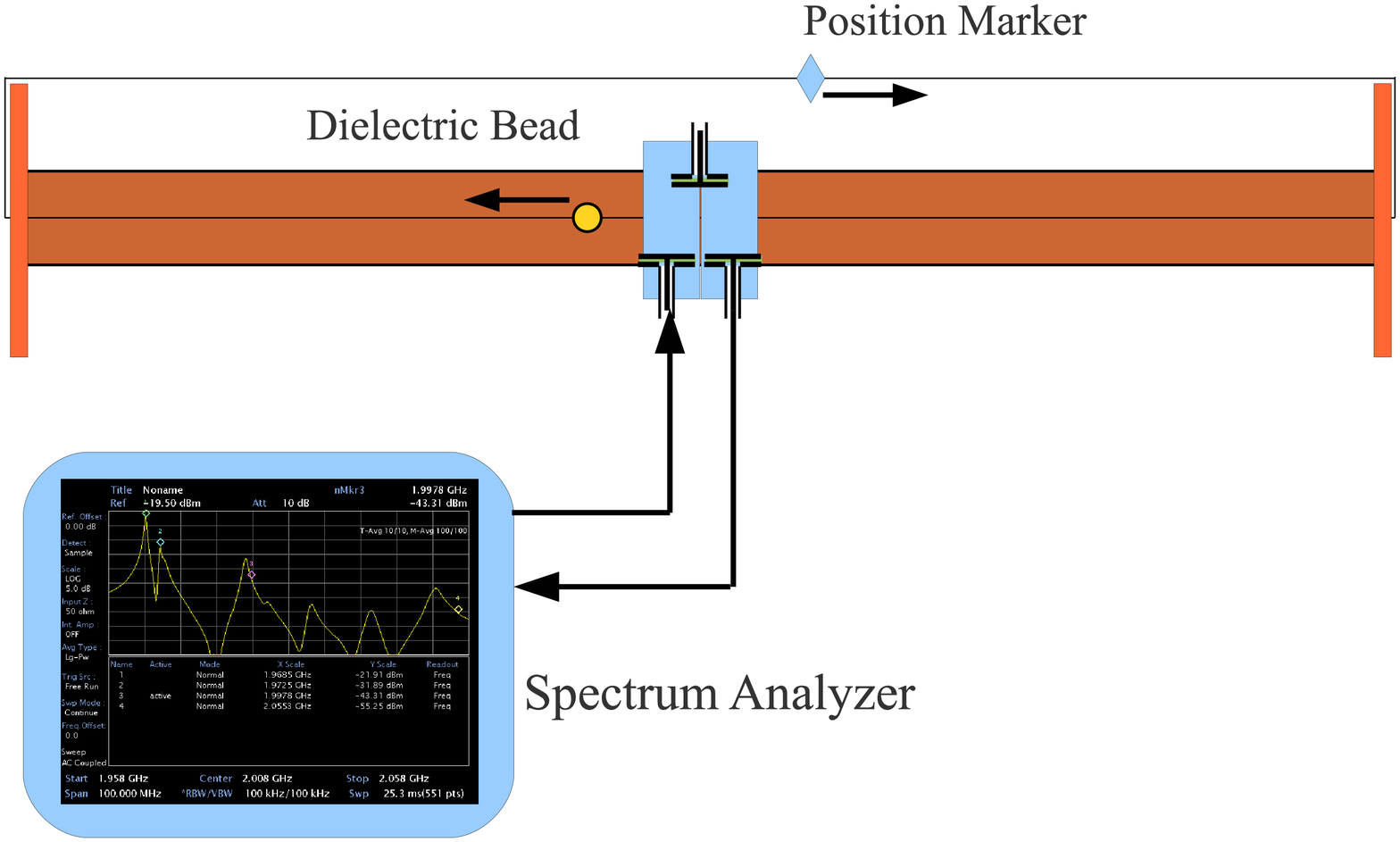}
   \caption{In the basic configuration of a bead pull measurement, a thin line is used to position the bead and a spectrum analyzer used to measure changes in the resonant frequencies.}
   \label{EC12_sikora:bead_setup}
   \end{center}
   \end{minipage}

   \begin{minipage}[t]{.48\textwidth}
   \begin{center}
   \includegraphics*[width=65mm]{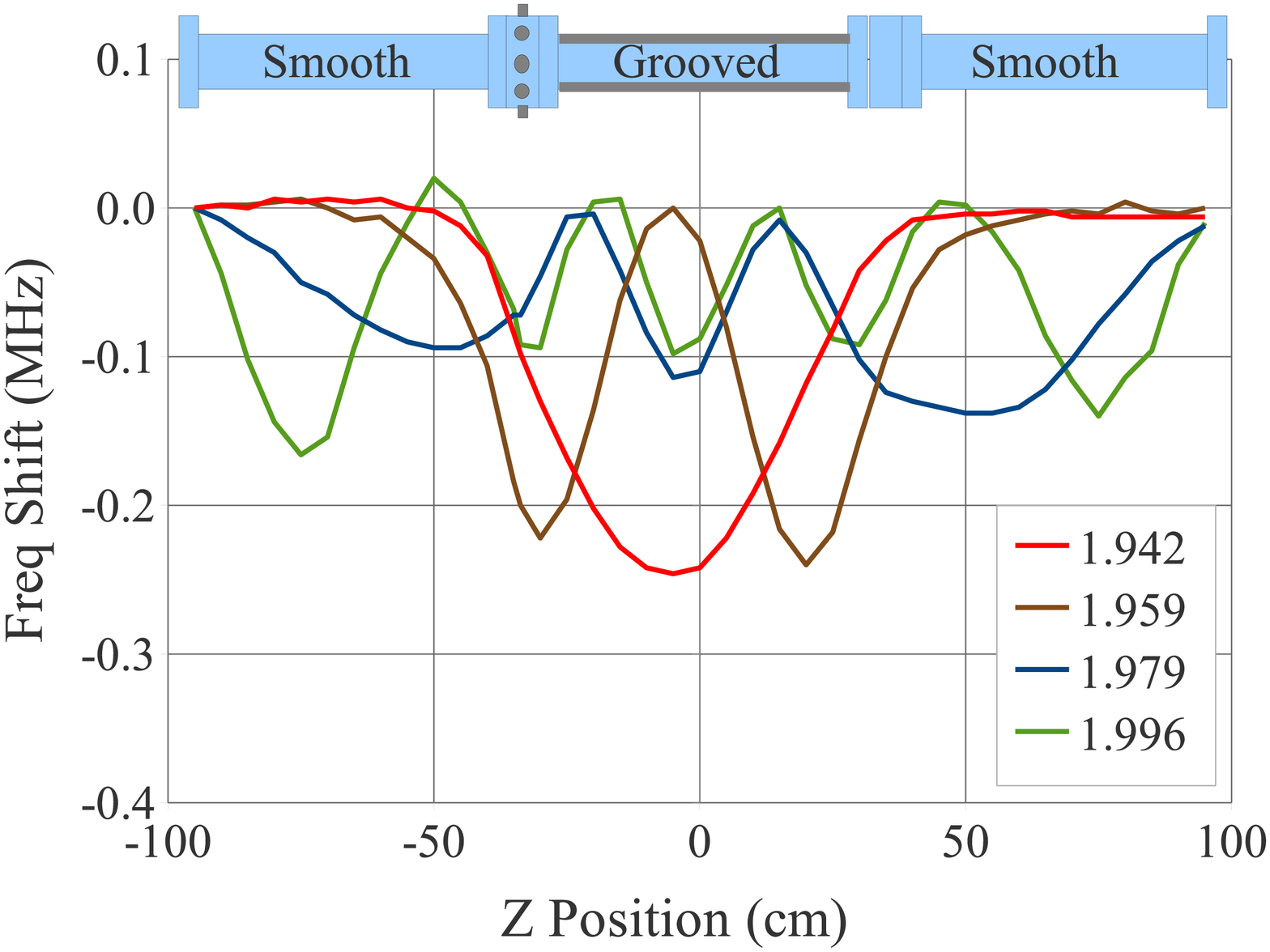}
   \caption{ In a bead pull measurement with grooved beam-pipe between two sections of smooth beam-pipe, the first two resonances are mostly confined to the grooved section of the chamber. Resonant frequencies are in GHz. 
   }
   \label{EC12_sikora:bead_sgs_h}
   \end{center}
   \end{minipage}

   \begin{minipage}[t]{.48\textwidth}
   \centering
   \includegraphics*[width=65mm]{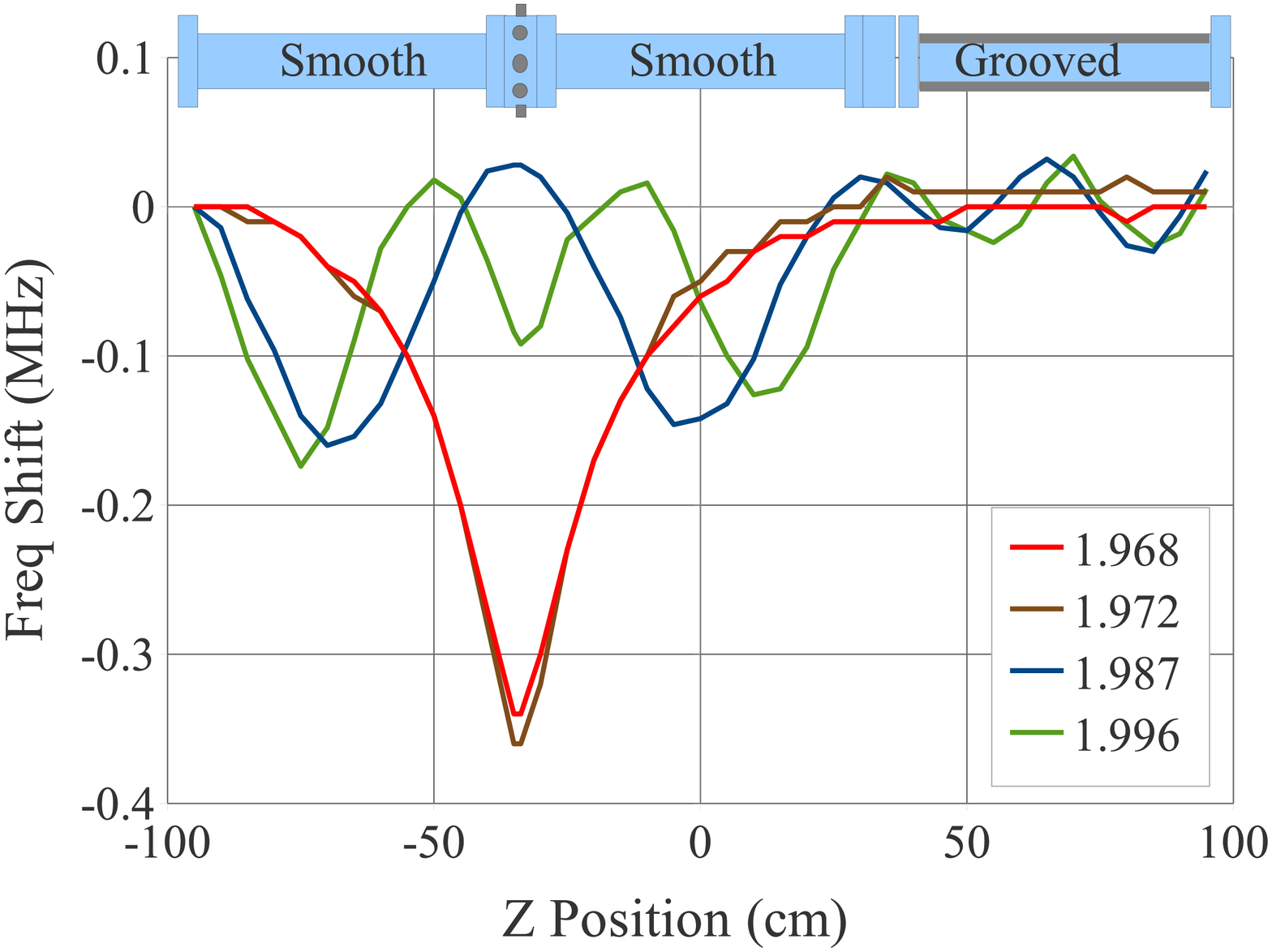}
   \caption{ In a bead pull measurement with smooth chambers on either side of the drive point, the first two are cutoff resonances that have an exponential decrease in field with distance from the drive point.}
   \label{EC12_sikora:bead_ssg_h}
   \end{minipage}
\end{figure}

Bead pull measurements have been widely used to measure the fields inside a resonant cavity~\cite{JAP23:Slater}. A dielectric bead positioned in a cavity will produce a shift in its resonant frequency that is proportional to $E^2$ at the location of the bead. This perturbation is analogous to Eq.~\ref{EC12_sikora:Ne}, except for a dielectric rather than a plasma. In performing a bead pull in resonant waveguide, the bead was attached to a thin mono-filament line that could be used to position the bead along the length of the waveguide as shown in Fig.~\ref{EC12_sikora:bead_setup}.  

Bead pull measurements were made using segments of round beam-pipe with geometries similar to the chambers in the L3 section of C{\small ESR}TA. There are two cross sections: the first is a smooth round chamber, the other extrusion has triangular grooves at the top and bottom of the chamber that are intended as an EC mitigation technique in dipole fields. 

Fig.~\ref{EC12_sikora:bead_sgs_h} shows the assembly of three sections of chamber along with their corresponding bead pull measurements. In this case, a grooved chamber was placed between two smooth chambers with conducting plates at either end of the assembly. The bead pull measurement shows that the fields of the first two resonances are predominantly within the grooved chamber only. This is due to the fact that the cutoff frequency of the grooved pipe is below that of the smooth pipe.  

Another bead pull measurement was made on an assembly having smooth beam-pipe on either side of the BPM drive point as shown in Fig.~\ref{EC12_sikora:bead_ssg_h}. The field appears to decrease exponentially with distance from the drive point. In the design of BPM flange, the buttons were recessed from the nominal inner diameter of the flange to prevent direct synchrotron radiation from striking the buttons when installed in the storage ring. The side effect of this modification is to lower the cutoff frequency in the small volume of the drive flange. This presumably explains the observed field pattern. 

\section{SIMULATIONS}

\subsection{Resonant TE Waves}

\begin{figure}[ht]
   \centering
   \includegraphics*[width=65mm]{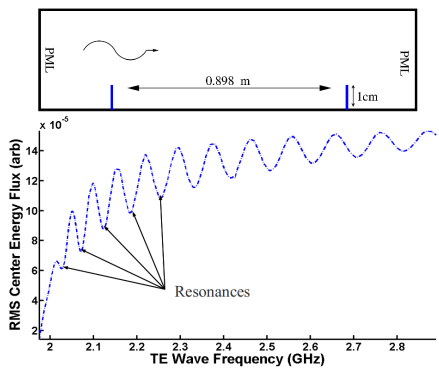}
   \caption{ In this simulation of a waveguide with two protrusions, resonances occur at the Poynting vector minima. The perfectly matched layers (PML) in the simulation serve to minimize reflections from the ends of the waveguide.
   }
   \label{EC12_sikora:poynting}
\end{figure}

The VORPAL~\cite{VORPAL}  simulation program was used to model standing waves in a waveguide. This was done by adding protrusions to the waveguide and observing the simulated response as shown in Fig.~\ref{EC12_sikora:poynting}. Resonant frequencies were determined by plotting the value of the simulated Poynting vector in the region between the protrusions. At a resonance, the sum of the Poynting vectors of the forward and reflected waves will be at a minimum. The resulting series of resonant frequencies is consistent with that of a rectangular cavity. In addition, when electron cloud was added into the simulation, the resonant frequencies shifted upward as expected from Eq.~\ref{EC12_sikora:Ne}.

\subsection{TE Wave Transmission in Magnetic Fields}

A magnetic field can change the magnitude of the resonant frequency shift for the same EC density
aside from the effect that the field might have on the cloud itself. This is especially true in the case of x-wave propagation, where the $E$ of the TE wave is perpendicular to the magnetic field. 

\begin{figure}[t]
   \centering
   \includegraphics*[width=65mm]{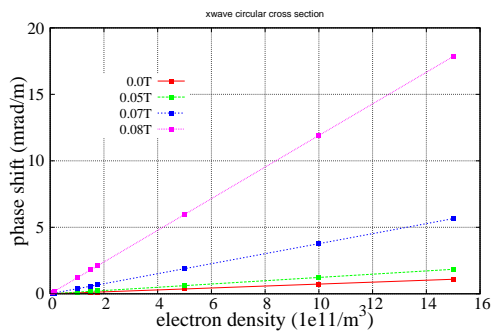}
   \caption{ This VORPAL simulation shows the change in phase shift vs. EC density for x-wave propagation for different magnetic fields. Fig.~\ref{EC12_sikora:hybrid} shows that this magnification is due to the proximity of the upper hybrid resonance.
   }
   \label{EC12_sikora:xwave}
\end{figure}

\begin{figure}[t]
   \centering
   \includegraphics*[width=65mm]{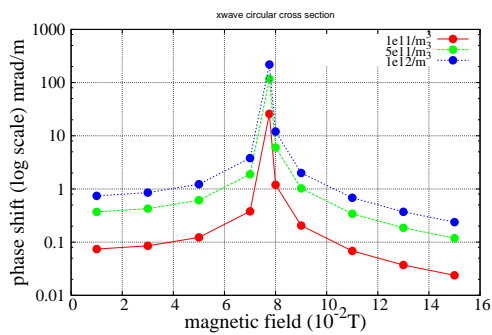}
   \caption{ In a simulation with a TE wave drive frequency of 2.17~GHz in circular waveguide, the upper hybrid resonance gives a substantial magnification of the phase shift of x-waves. At this frequency, the upper hybrid resonance would occur at $7.757\times10^{-2}$~T.
   }
   \label{EC12_sikora:hybrid}
\end{figure}

Simulations using VORPAL have shown the effect of a magnetic field on the phase shift of x-waves propagating through a circular waveguide. At low magnetic fields (near the upper hybrid resonance) the effect is a magnification of the phase shift per unit length. The degree of magnification is determined by the proximity to the upper hybrid resonance where the excitation frequency is the sum of the cyclotron and plasma frequencies $ \omega_{cyclotron} + \omega_p$. Fig.~\ref{EC12_sikora:xwave} shows an increasing change in phase shift at different magnetic fields. A more general view of this is given in Fig.~\ref{EC12_sikora:hybrid} where the phase shift is plotted vs. the magnetic field at a given excitation frequency.

\begin{figure}[ht]
   \centering
   \includegraphics*[width=65mm]{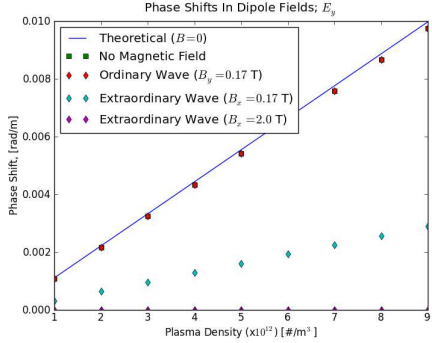}
   \caption{ This simulation of propagation through an electron cloud in a dipole field shows the phase shift of the o-wave little affected by the magnetic field, while the x-wave phase shift is close to zero at 2~T.
   }
   \label{EC12_sikora:veitzer_xo}
\end{figure}

Simulations were also run for TE waves propagating in high magnetic fields (well above the upper hybrid resonance). Fig.~\ref{EC12_sikora:veitzer_xo} shows a VORPAL simulation of the propagation of 1.66~GHz TE waves through an electron cloud in a dipole field. The phase shift of the o-wave is little affected by the magnetic field, while the x-wave phase shift is close to zero at 2~T. In a high magnetic field that is perpendicular to the $E$ field, the electrons are no longer free to move in that plane. In contrast the o-waves, with the magnetic field parallel to the $E$ field, have phase shifts that are little affected by magnetic fields up to 2~T. But in circular waveguide, the transverse field lines are curved and will have some component perpendicular to the magnetic field. So some reduction in the phase shift should be expected as the magnetic fields become very large.

\begin{figure}[ht]
   \centering
   \includegraphics*[width=65mm]{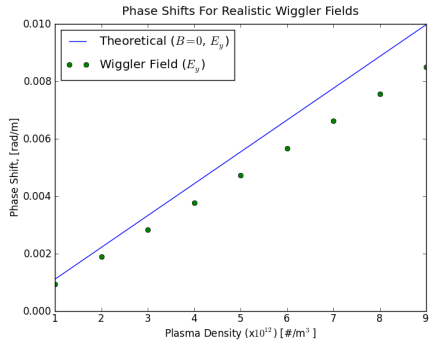}
   \caption{ Simulation of the phase shift in a wiggler field shows a reduction in the phase shift even when the $E$ field is parallel to the B field.
   }
   \label{EC12_sikora:veitzer_wiggler}
\end{figure}

Fig.~\ref{EC12_sikora:veitzer_wiggler} is a simulation of the phase shift in a wiggler field. The waveguide is rectangular so that the $E$ field is parallel to the nominal wiggler field. Unlike the simulation in a dipole field, the wiggler field gives phase shift that is somewhat reduced from the zero field value~\cite{IPAC12:Veitzer}. This simulation result needs further study.  

\section{Data with Magnetic Field}

\begin{figure}[t]
   \centering
   \includegraphics*[width=65mm]{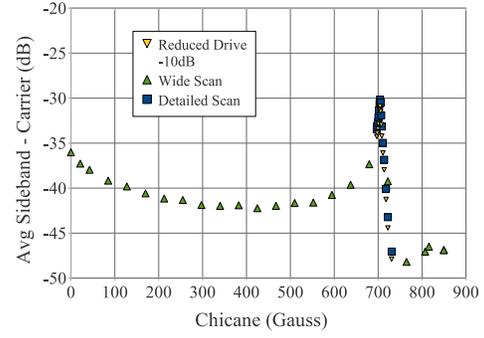}
   \caption{ With x-wave excitation of 1.9714~GHz, data was taken with a dipole field ranging from 0 to about 850 Gauss. The sideband amplitude greatly increases at a field that corresponds to the upper hybrid resonance. The beam current was 140~mA of positrons in 20 bunches.  
   }
   \label{EC12_sikora:chicane}
\end{figure}

A resonant TE wave measurement was made in a dipole field at C{\small ESR}TA with the result shown in 
Fig.~\ref{EC12_sikora:chicane}. An x-wave was excited in the round beam-pipe at 1.9714~GHz and the dipole field ramped from 0 to about 850~Gauss. The dipole is part of a four magnet chicane. The beam current in the storage ring was 140~mA in 20 bunches of positrons spaced at 14ns. The data clearly shows that the upper hybrid resonance magnifies the sideband amplitudes as might be expected from the simulation the phase shift of Fig.~\ref{EC12_sikora:hybrid}. However a retarding field analyzer (RFA), that measures the electron cloud current in the same chamber, also shows an increased signal at this magnetic field. Further, this increase in the RFA signal goes away when the TE wave excitation is turned off. At the moment, we do not know to what extent the excitation of the TE wave changes the response of the RFA or produces an increase the electron cloud in the chamber.

\section{CONCLUSIONS AND FUTURE WORK}

In many cases, the excitation of TE waves in beam-pipe is dominated by resonances rather than point to point transmission. Modeling the beam-pipe as a resonant cavity results in a simple expression for the shift in resonance vs. EC density. The effect that this frequency shift has on the observed signals is complicated by the fact that the damping time of the resonance is similar to the electron cloud duration. So a full treatment of the problem will involve the effect of the transient response of the resonance to a rapidly varying EC density. Approximate values for the EC density can be obtained by taking the duration of the cloud to be long compared to the damping time of the resonance. 

Bead pull measurements can be an aid in the understanding of the resonant fields. Bench measurements show that localized measurement of the EC density is possible in some geometries.

Simulations have shown the effects of dipole magnetic fields on the phase shift of TE waves propagating through an electron cloud. At low magnetic fields there is an enhancement of this phase shift depending upon proximity to the upper hybrid resonance. At high fields, there is significant suppression of the phase shift for x-waves propagating through the EC density, while the o-waves seem unaffected. 

Future efforts will include the application of the results of both bench tests and simulations to measurements made in accelerators. It is important to obtain both the magnitude of the EC density measurement and understand \textit{where} the measurement is being made. This is especially true where specialized chambers or devices such as clearing electrodes or grooved chambers have been installed.

\section{ACKNOWLEDGEMENTS}

We would like to thank to Seth Veitzer at Tech-X Corp. and the VORPAL Development Team  for their help with VORPAL simulations. We are grateful for the support of Yulin Li and the vacuum group at Cornell in setting up bead pull measurements on beam-pipe. We would also like to acknowledge the significant contributions of past undergraduates Ben Carlson and Ken Hammond who were part of the Research Experience for Undergraduates Program (REU) of the U.S. National Science Foundation.


\end{document}